%% file: main.tex
\documentclass[a4paper,11pt]{article}
\usepackage{pos}

\usepackage{amsmath}
\usepackage{siunitx}
\usepackage{physics}
\usepackage{dsfont}
\usepackage{subcaption}
\usepackage{multirow}
\usepackage{cleveref}

\ExplSyntaxOn
\msg_redirect_name:nnn { siunitx } { physics-pkg } { none }
\ExplSyntaxOff

\newcommand{\eff}{\text{eff}}
\newcommand{\zth}{^{(0)}}
\newcommand{\fst}{^{(1)}}
\newcommand{\qcdiso}{$\text{QCD}_{\text{iso}}$}

\newcommand{\pr}{\text{pr}}

\title{Precision Determination of Baryon Masses including Isospin-breaking}

\author*[a]{Alexander M. Segner}
\author[b,c]{Andreas Risch}
\author[a,d,e]{Hartmut Wittig}

\affiliation[a]{Institut für Kernphysik, Johannes Gutenberg-Universität, \\
    Mainz, Johann-Joachim-Becher-Weg 45, 55128 Mainz, Germany}

\affiliation[b]{Department of Physics, University of Wuppertal, \\
    Gaussstr. 20, 42119 Wuppertal, Germany}

\affiliation[c]{John von Neumann-Institut für Computing NIC, Deutsches Elektronen-Synchrotron DESY,
    \\
    Platanenallee 6, 15738 Zeuthen, Germany}

\affiliation[d]{Helmholtz Institut Mainz, \\
    Staudingerweg 18, 55128 Mainz, Germany}

\affiliation[e]{Helmholtzzentrum für Schwerionenforschung, \\
    64291 Darmstadt, Germany}

\emailAdd{alsegner@uni-mainz.de}

\abstract{
    We give an update on an ongoing project in which we calculate the masses of octet and decuplet
    baryons including isospin-breaking effects.
    To this end, we employ single- and two-state-fits to effective masses up to leading order in the
    expansion in isospin-breaking parameters.
    In order to remove subjective bias on asymptotic masses we furthermore compute an AIC-based
    model-average of our fits, for which we show results on ensembles at lattice spacings of 0.064
    fm and 0.076 fm with corresponding pion masses ranging from 220 MeV to 360 MeV.
}

\FullConference{The 40th International Symposium on Lattice Field Theory (Lattice 2023)\\
July 31st - August 4th, 2023\\
Fermi National Accelerator Laboratory\\}

\begin{document}
\maketitle

\input{introduction}

\input{setup}

\input{ib}

\input{methods}

\input{results}

\input{conclusion}

\input{acknowledgements}

\setlength{\bibsep}{0pt}
\bibliographystyle{JHEP}
\bibliography{bibliography.bib}

\end{document}

%% file: introduction.tex
\section{Introduction}

As the precision of lattice QCD calculations improves, effects stemming from QED and strong
isospin-breaking need to be accounted for to meet the precision targets.
This is particularly necessary for observables such as the hadronic contributions to the anomalous
magnetic moment of the muon, $(g-2)_\mu$, as for example computed in \cite{Ce:2022kxy}.
This quantity's uncertainty is strongly influenced by the lattice scale \cite{DellaMorte:2017dyu}
whose value needs to be determined at the per-mil level to be competitive with the direct
measurement.
For the isospin-symmetric $N_f=2+1$ CLS ensembles \cite{Bruno:2014jqa}, this goal can only be
reached by incorporating isospin-breaking effects into the determination of the lattice scale.

Thus far, the scale for the CLS ensembles has been computed from a combination of pion and kaon
decay constants \cite{Bruno:2016plf,Strassberger:2021tsu}, which yield very precise results, but the
introduction of isospin-breaking corrections for these observables proves conceptually difficult on
the lattice \cite{Carrasco:2015xwa}.
As a tradeoff between complexity and overall precision, we investigate the viability of using the
masses of the lowest-lying baryon octet and decuplet states as alternative scale setting quantities,
since isospin-breaking corrections for these can be computed reliably using a perturbative approach
introduced by the RM123 collaboration \cite{deDivitiis:2011eh,deDivitiis:2013xla}.

We give a short overview of our simulation setup and the baryonic operators we use in
\cref{sec:setup}, followed by a brief introduction to the expansion of the isospin-broken theory
around the isospin-symmetric one in \cref{sec:ib}.
Afterwards, we explain the methodology we use to extract the ground state masses of the different
baryonic states in \cref{sec:methods} and finally summarize our results in \cref{sec:results} before
concluding in \cref{sec:conclusion}.

%% file: setup.tex
\section{Simulation Setup} \label{sec:setup}

For this project, we use ensembles generated by the \textit{Coordinated Lattice Simulations} (CLS)
effort \cite{Bruno:2014jqa} with $N_f=2+1$ flavours of non-perturbatively $\mathcal O(a)$-improved
Wilson fermions \cite{Luscher:1996sc} and a tree-level Lüscher-Weisz gauge action
\cite{Bulava:2013cta}.
We apply APE smearing \cite{APE:1987ehd} to the QCD gauge links and for the quark sources we use
$\text{SU}(3)$-covariantly Wuppertal-smeared \cite{Gusken:1989qx} point sources.
The smearing parameters are tuned such that the smearing radius \cite{UKQCD:1993gym} is
approximately \SI{0.5}{fm} and that the nucleon effective mass is minimized at an early time on the
H105 ensemble.

We compute correlators for all octet and decuplet baryons using a subset of interpolating operators
introduced by the Lattice Hadron Physics Collaboration \cite{Basak:2005ir} which we found to have
the best overlap with their respective ground states.
The interpolators we use are listed in \cref{tab:operators}.

\begin{table}[htb]
    \centering
    \caption{Baryonic operators used for the various states considered in this project.
    The operators in each row describe the same state for different spin-$z$ components (arranged in
    descending order) and the negative-parity-operators are the parity partners of the respective
    positive-parity-operators.
    For the conventions used in this notation, we refer to \cite{Basak:2005ir}.}
    \input{tables/operatortable.tex}
    \label{tab:operators}
\end{table}

The two-point-functions from the different operators in one row of \cref{tab:operators} are averaged
and then combined with the time-reversed two-point-functions of the opposite parity state to reduce
noise.
Furthermore, we increase the available statistics using the truncated solver method
\cite{Blum:2012my,Blum:2012uh,Shintani:2014vja} with 32 sources per gauge-configuration and one
source for bias-correction, reducing the computational cost of inversions.

%% file: tables/operatortable.tex
\begin{tabular}{r|c|l}
    Baryon & Parity & Operators \\
    \hline\hline
    \multirow{2}{*}{$N/\Lambda$} & g & $\sqrt{2} N_{121}$, $\sqrt{2} N_{122}$ \\
    \cline{2-3}
    & u & $\sqrt{2} N_{343}$, $\sqrt{2} N_{344}$ \\
    \hline
    \multirow{2}{*}{$\Sigma/\Xi$} & g & $\sqrt{\frac23}\left(\Sigma_{112} - \Sigma_{121}\right)$, $\sqrt{\frac23}\left(\Sigma_{122} - \Sigma_{221}\right)$ \\
    \cline{2-3}
    & u & $\sqrt{\frac23}\left(\Sigma_{334} - \Sigma_{343}\right)$, $\sqrt{\frac23}\left(\Sigma_{344} - \Sigma_{443}\right)$ \\
    \hline
    \multirow{2}{*}{$\Delta/\Omega$} & g & $\Delta_{111}$, $\sqrt{3} \Delta_{112}$, $\sqrt{3} \Delta_{122}$, $\Delta_{222}$ \\
    \cline{2-3}
    & u & $\Delta_{333}$, $\sqrt{3} \Delta_{334}$, $\sqrt{3} \Delta_{344}$, $\Delta_{444}$ \\
    \hline
    \multirow{2}{*}{$\Sigma^*/\Xi^*$} & g & $\Sigma_{111}$, $\frac1{\sqrt{3}}\left(\Sigma_{112} + 2 \Sigma_{121}\right)$, $\frac1{\sqrt{3}}\left(2 \Sigma_{122} + \Sigma_{221}\right)$, $\Sigma_{222}$ \\
    \cline{2-3}
    & u & $\Sigma_{333}$, $\frac1{\sqrt{3}}\left(\Sigma_{334} + 2 \Sigma_{343}\right)$, $\frac1{\sqrt{3}}\left(2 \Sigma_{344} + \Sigma_{443}\right)$, $\Sigma_{444}$ \\
    \hline
\end{tabular}

%% file: ib.tex
\section{Expansion in Isospin-Breaking Parameters}\label{sec:ib}

We compute isospin-breaking corrections to correlation functions $C(t)=\ev{\mathcal
B(t)\overline{\mathcal B}(0)}$ with baryonic operators $\mathcal B$ as described in \cref{sec:setup}
using an expansion of full QCD+QED about the isospin-symmetric theory \qcdiso\ in a manner first
introduced by the RM123 collaboration \cite{deDivitiis:2011eh,deDivitiis:2013xla}.
This expansion in terms of the electromagnetic coupling $e$ and the differences in quark masses
$\Delta m_f$ between QCD+QED and \qcdiso\ for $f\in\{u,d,s\}$ is given by
\[
    C^\varepsilon(t)=C^{\varepsilon\zth}(t)
        +\sum_f\Delta m_f\pdv{C^\varepsilon(t)}{m_f}\eval_{\varepsilon=\varepsilon\zth}
        +e^2\pdv{C^\varepsilon(t)}{e^2}\eval_{\varepsilon=\varepsilon\zth}
        +\mathcal O(\Delta\varepsilon^2).
\]
Here, the superscripts $\varepsilon$ and $\varepsilon\zth$ indicate whether an expression is
evaluated in QCD+QED or \qcdiso\ respectively.
Diagramatically, the above expansion, disregarding quark-disconnected contributions, is
\begin{align*}
    \ev{\vcenter{\hbox{\includegraphics[width=2.8cm]{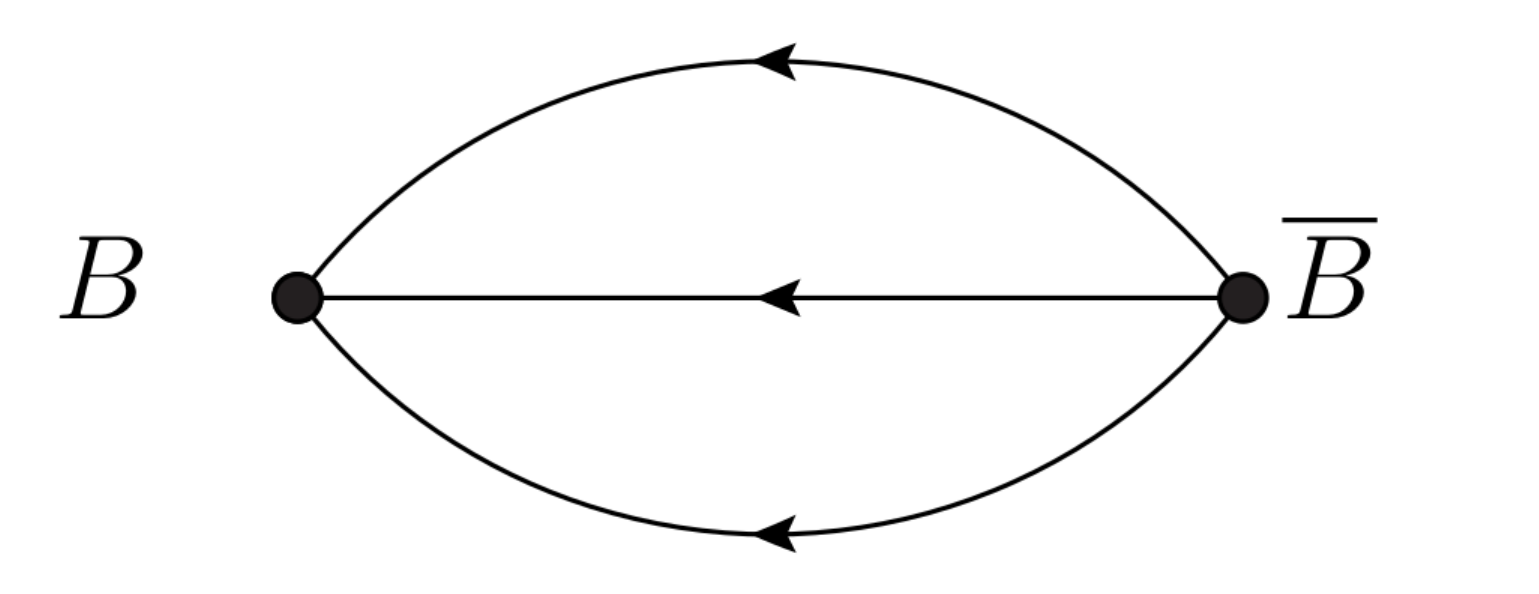}}}}^{\varepsilon}=&
        \ev{\vcenter{\hbox{\includegraphics[width=2.8cm]{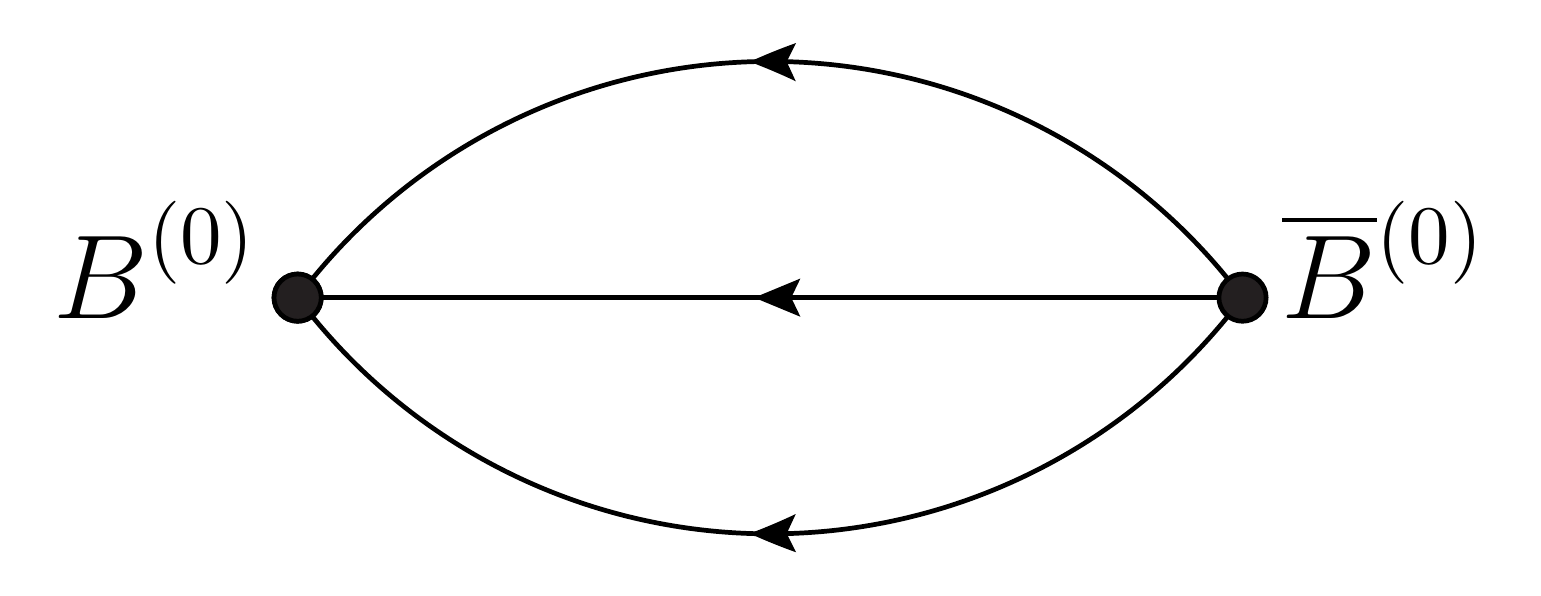}}}}^{\varepsilon^{(0)}}+
        \sum_f\Delta m_f\ev{\vcenter{\hbox{\includegraphics[width=2.8cm]{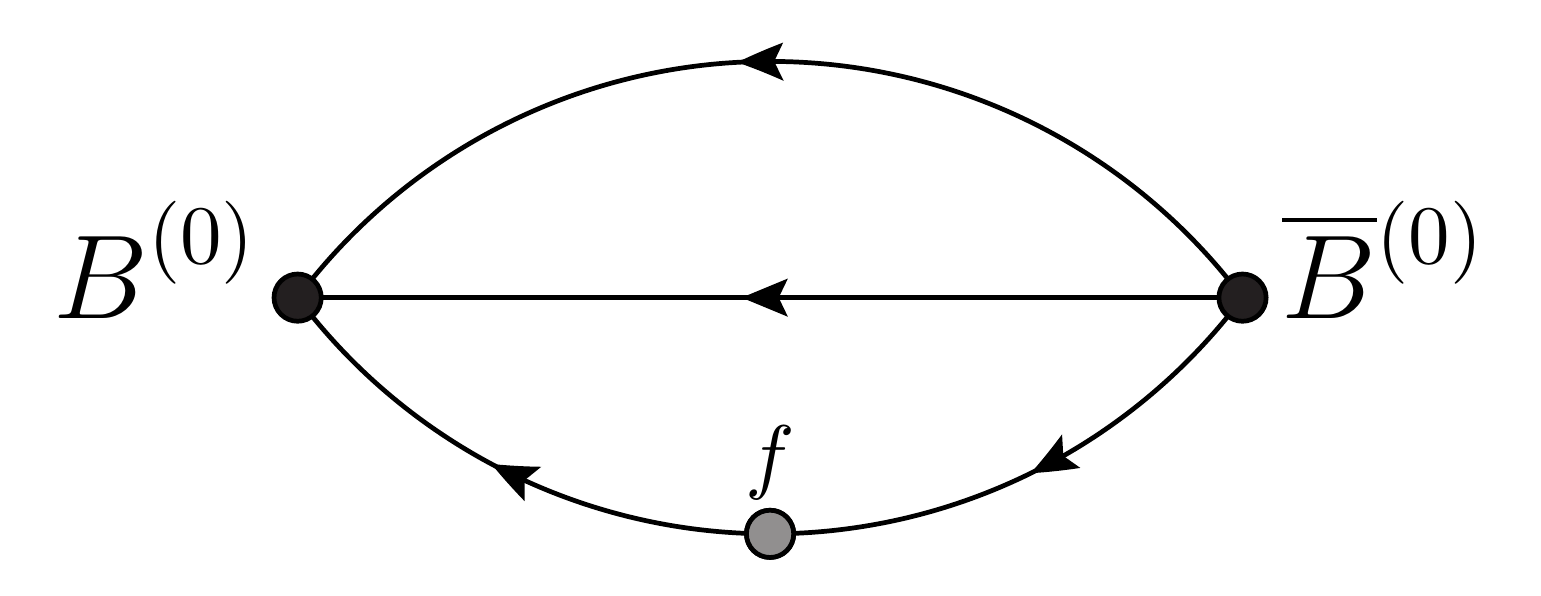}}}}^{\varepsilon^{(0)}}
        \\
        &+ e^2\ev{\vcenter{\hbox{\includegraphics[width=2.8cm]{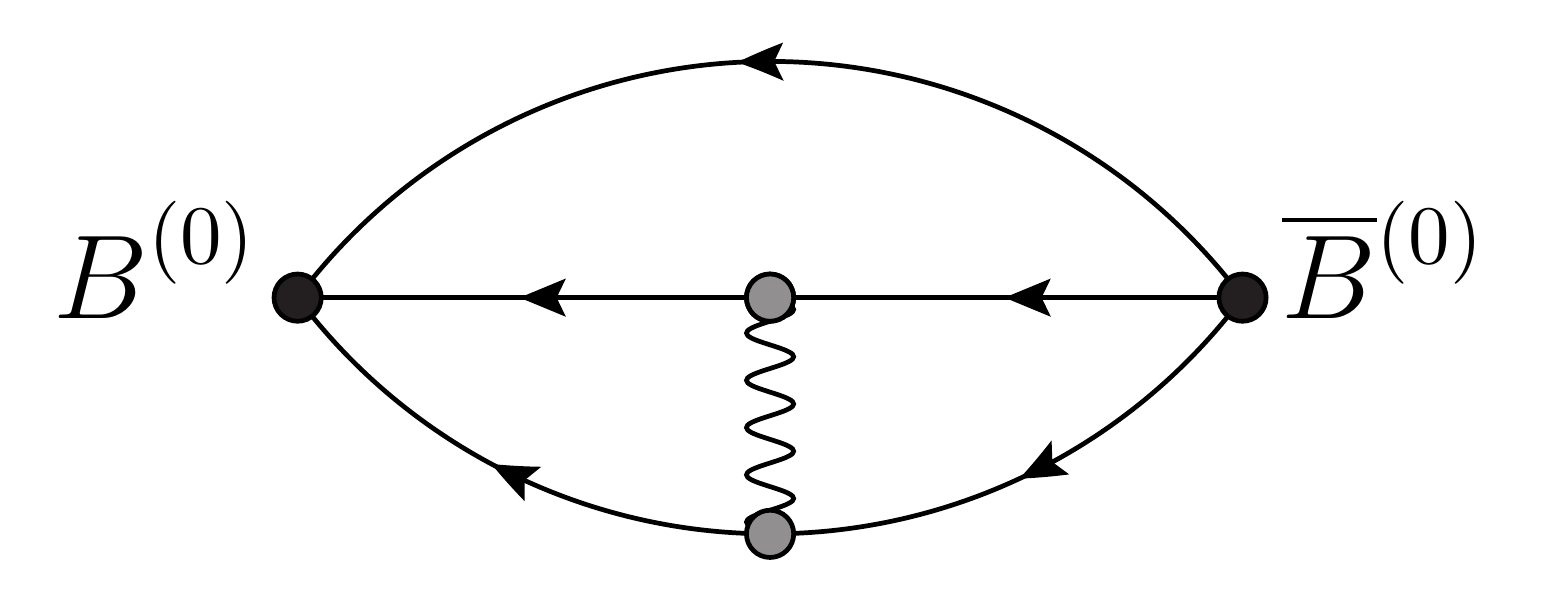}}} +
        \vcenter{\hbox{\includegraphics[width=2.8cm]{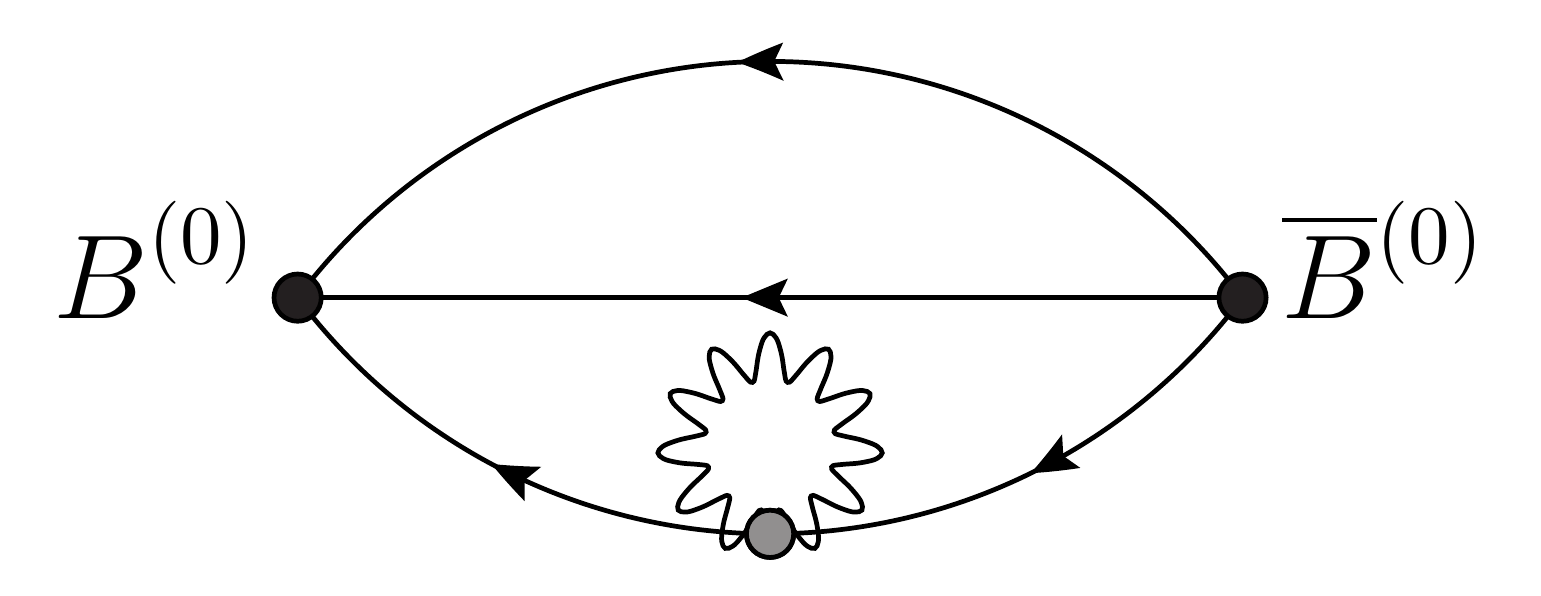}}} +
        \vcenter{\hbox{\includegraphics[width=2.8cm]{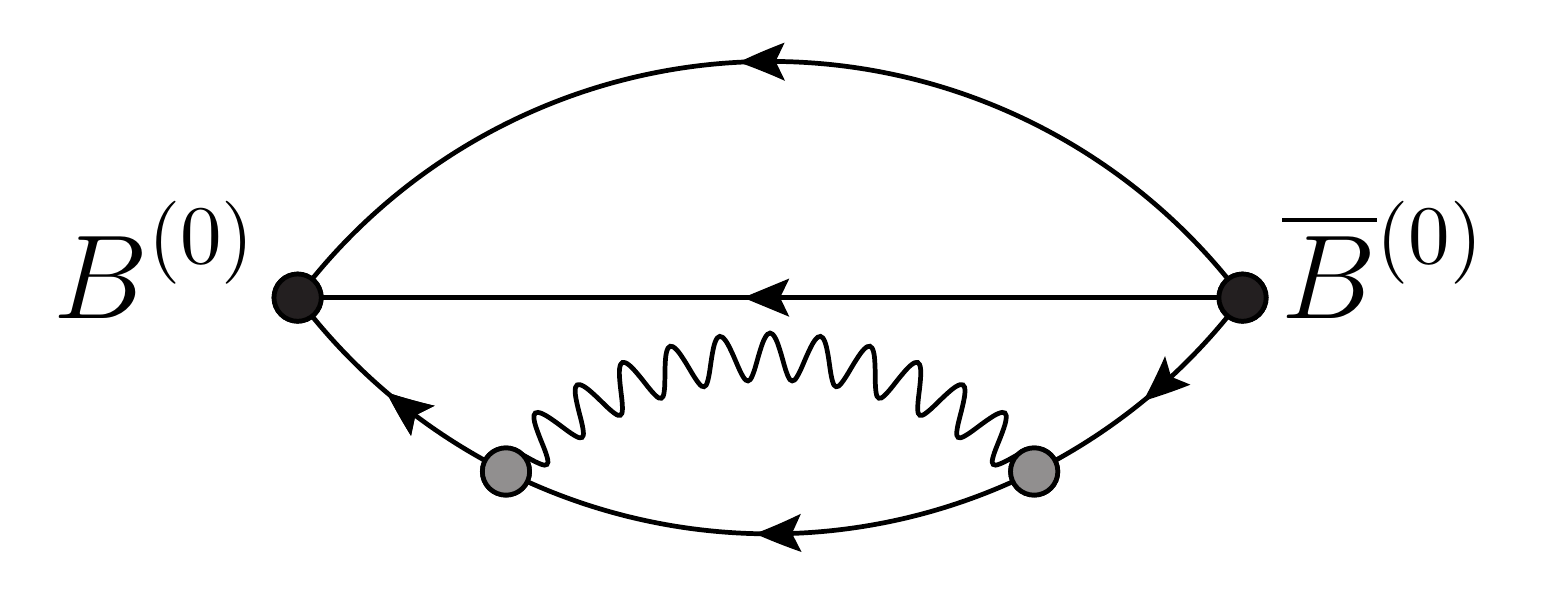}}}}^{\varepsilon^{(0)}}
\end{align*}
where $B$ encodes the colour-, spin-, and flavour-structure of $\mathcal B$.
Thus far, we do not consider sea-quark interactions in this expansion.
However, we do calculate all possible diagrams in which a photon line is connected to one of the
quarks with the other end left open.
Hence, we can use these diagrams in combination of an equivalent diagram of a quark loop with a
photon vertex should we decide to investigate these contributions at a later stage.
The propagators including the isospin-breaking corrections are computed using sequential propagators
with the corresponding operator insertions.
For the computation of the QED corrections, we use the $\text{QED}_L$ prescription
\cite{Hayakawa:2008an} in Coulomb gauge \cite{Risch:2018ozp,Risch:2021nfs,Risch:2021hty}.

%% file: methods.tex
\section{Analysis Methods}\label{sec:methods}

The calculation of baryon masses in our setup is based on effective masses in the isospin-symmetric
and LO isospin-breaking contributions to the correlation functions.
The definitions of these effective masses are motivated by the asymptotic functional behaviour of
a simple two-point function $C(t)=ce^{-mt}$ and its expansion in terms of isospin-breaking
coefficients
\[
    C_i\fst(t)=\qty(c_i\fst-c\zth m_i\fst t)e^{-m\zth t},
\]
where each quantity is expanded as $X=X\zth+\sum_i\Delta\varepsilon_iX_i\fst+\mathcal
O(\Delta\varepsilon^2)$ for $X\in\{C,m,c\}$ with $\Delta\varepsilon=(\Delta m_u,\Delta m_d, \Delta
m_s, e^2)$.
In these asymptotic forms, the ground state masses can be calculated as \cite{Segner:2022dou}
\begin{align}
    \begin{aligned}\label{eq:meff_continuum}
        m\zth=&-\dv t\log(C\zth(t))\text{\hspace{.6cm} and \hspace{.6cm}}
        m\fst=-\dv t\frac{C_i\fst(t)}{C\zth(t)},
    \end{aligned}
\end{align}
which can be computed on the lattice via the discretizations
\begin{align}
    \begin{aligned}\label{eq:meff_lattice}
        (am_\eff)\zth(t)=&\log\qty(\frac{C\zth(t)}{C\zth(t+a)})\text{\hspace{.6cm} and \hspace{.6cm}}
        (am_\eff)_i\fst(t)=\frac{C_i\fst(t)}{C\zth(t)}-\frac{C_i\fst(t+a)}{C\zth(t+a)}.
    \end{aligned}
\end{align}
While these definitions result in functions converging to plateaus for large $t$, in practice, the
baryon noise problem \cite{Lepage:1989hd,Wagman:2016bam} often hides these plateaus in the
exponentially growing noise, making it difficult to determine a reasonable fit interval for
single-state fits.
This leads us to incorporate two-state fit ansätze \cite{DelDebbio:2006cn} into our analysis, which
take the forms
\begin{align}
    (am)_\eff\zth(t)=&am\zth+\gamma e^{-\Delta M\zth t}\text{\hspace{.6cm} and}  \label{eq:fit0} \\
    (am)_{i,\eff}\fst(t)=&am_i\fst+\qty(\alpha_i-\beta_i t)e^{-\Delta M\zth t} \label{eq:fit1}
\end{align}
for the isospin-symmetric and the isospin-breaking contributions, respectively
\cite{Segner:2022dou}.

Note, that the parameter $\Delta M\zth$ is the same in all contributions.
We thus plug the values obtained from fits to \cref{eq:fit0} into \cref{eq:fit1} when fitting the
first order, which simplifies the fits in the isospin-breaking corrections to a point that the
$\chi^2$-minimization can be solved analytically.
In order to eliminate any bias in the choice of fit interval for a given fit type, we adopt a
model-averaging technique based on the \textit{Akaike information criterion} (AIC)
\cite{Akaike:1998zah,Jay:2020jkz,Neil:2022joj} for Gaussian noise.
This average is defined via expectation values of given fit parameters according to a distribution
on the space of fit models assigning each model $M$ a probability
\begin{align}
    \pr(M|D)\propto\exp(-\frac12\chi^2(M,D)-k(M)-n(M,D))\label{eq:AIC_weight}
\end{align}
given the fitted data $D$, where $k$ is the number of fit parameters of model $M$ and $n$ is the
number of data points in $D$ not considered in the fit.
The inclusion of the term $n$ is necessary as this allows for varying fit intervals which the AIC
does not account for in its usual form $\text{AIC}=\chi^2+2k$.

\begin{figure}[htb]
    \centering
    \begin{subfigure}[t]{0.6\textwidth}
        \includegraphics[width=\textwidth]{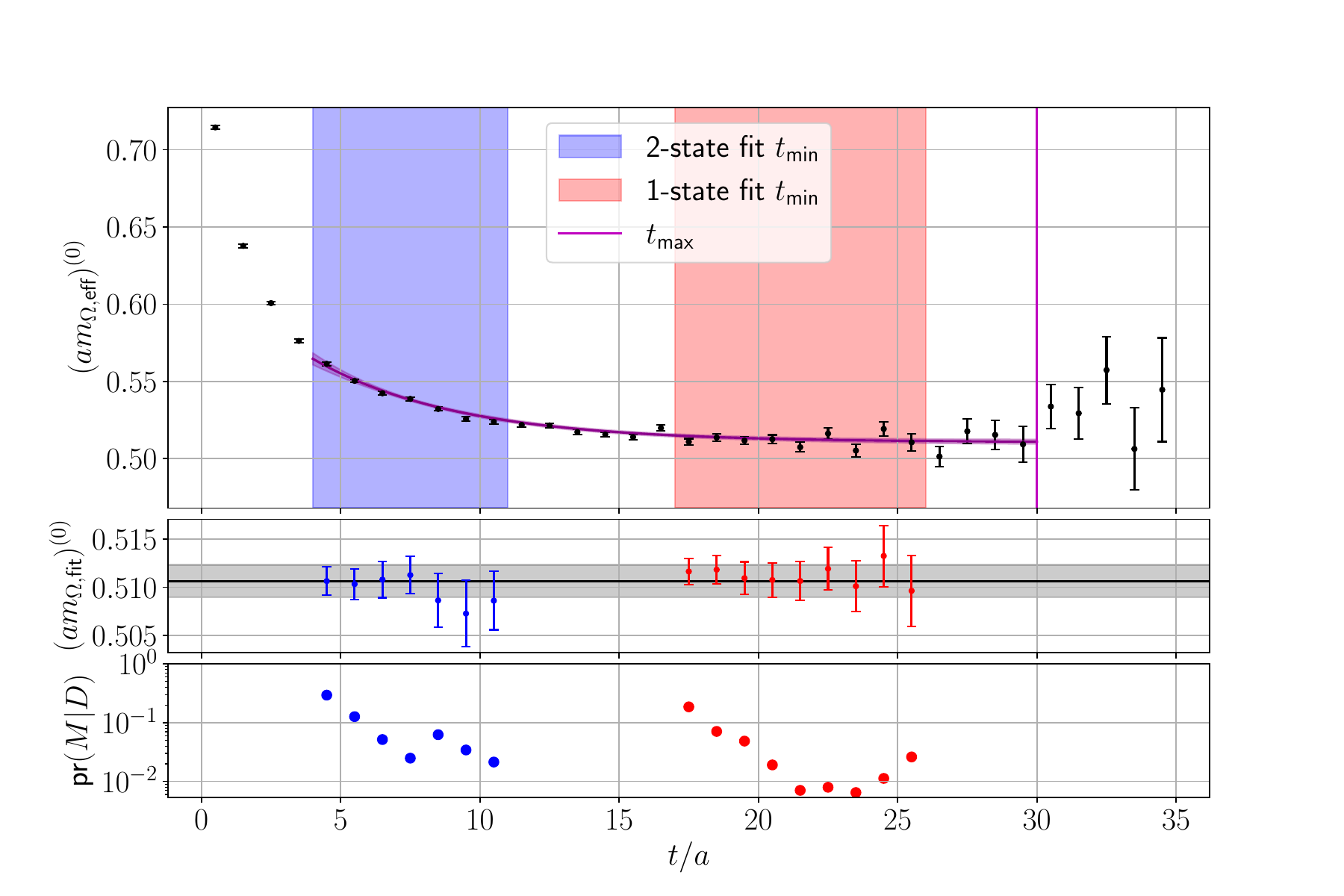}
        \subcaption{}
        \label{sfig:aic_omega_sym}
    \end{subfigure}
    \begin{subfigure}[t]{0.48\textwidth}
        \includegraphics[width=\textwidth]{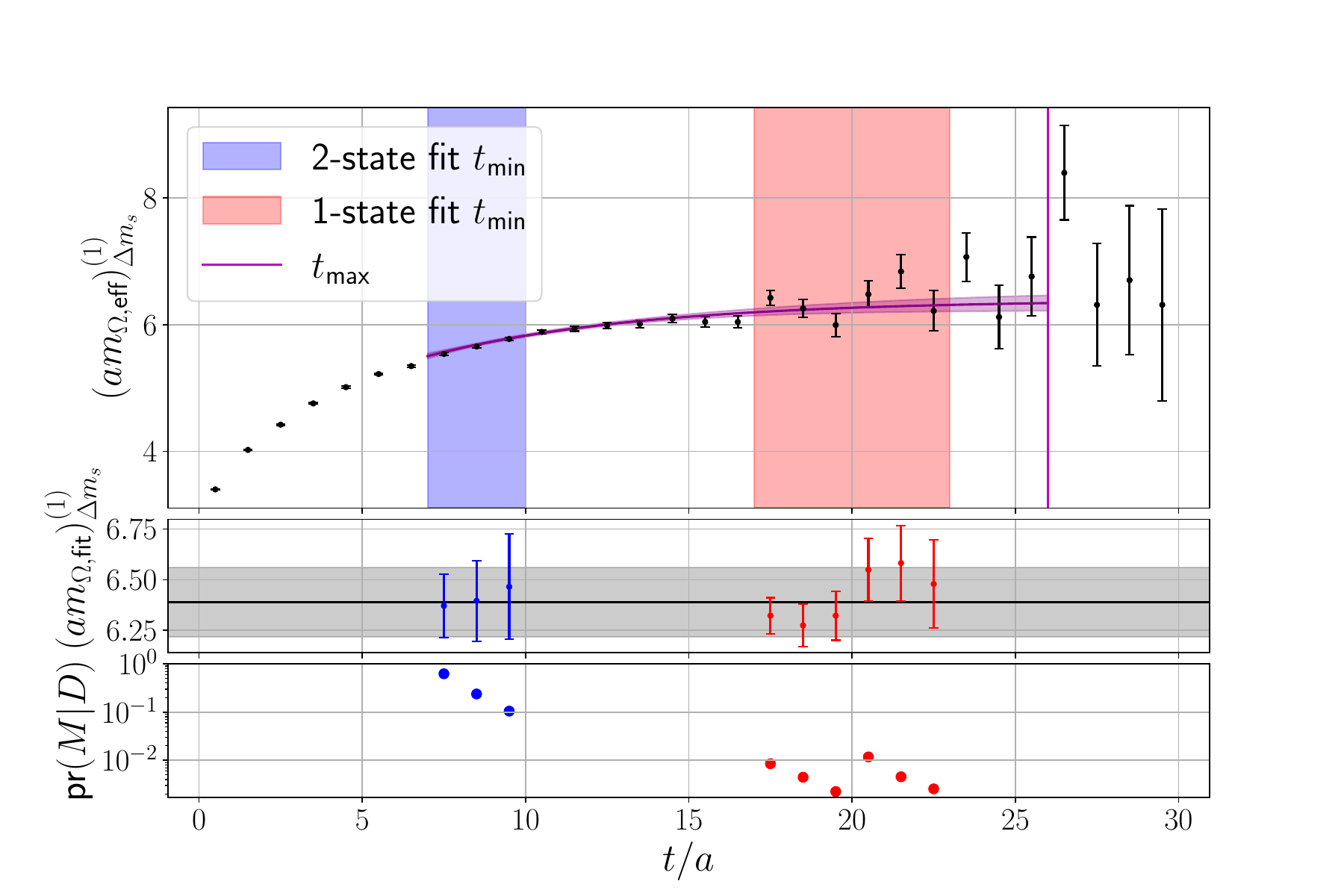}
        \subcaption{}
        \label{sfig:aic_omega_s}
    \end{subfigure}
    \begin{subfigure}[t]{0.48\textwidth}
        \includegraphics[width=\textwidth]{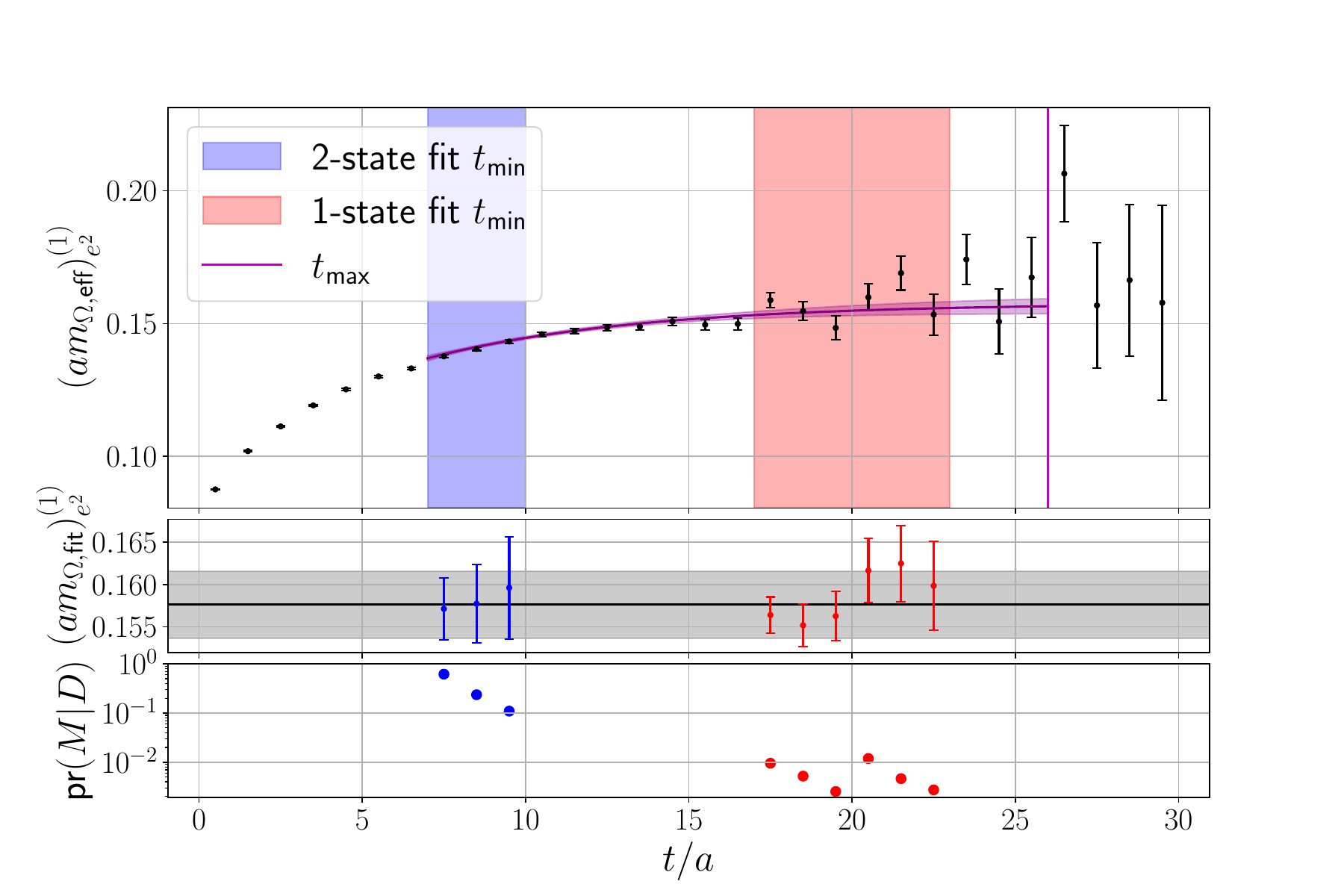}
        \subcaption{}
        \label{sfig:aic_omega_qed}
    \end{subfigure}
    \caption{
        Example plots for the AIC model averages for the $\Omega$ baryon on the N200 ensemble.
        The top panels show the effective masses as defined in \cref{eq:meff_lattice} for the
        isospin-symmetric contribution (a), the $\Delta m_s$ correction (b), and the electromagnetic
        corrections (c).
        The blue and red bands show the ranges from which the values of $t_{\text{min}}$ were
        chosen, respectively.
        The purple vertical line shows the end of the fit intervals, which is common to all fits.
        The central panels show the results for the respective asymptotic mass for the individual
        fits where each point belongs to the fit starting at the the time $t/a$ shown on the
        abscissa.
        The bottom panels show the model weight for the respective fits as defined in
        \cref{eq:AIC_weight}.
    }
    \label{fig:aic_example}
\end{figure}

If fit model $M_i$ predicts values $\ev{a_0}_{M_i}\in\mathds R^l$ with covariance matrix
$C_i\in\mathds R^{l\times l}$ for a set of fit parameters, the model average is defined as
\begin{align}
    \ev{a_0}=\sum_i\ev{a_0}_{M_i}\pr(M_i|D)\label{eq:AIC_mean}
\end{align}
and the resulting covariance matrix is given by
\begin{align}
    C=&\sum_i C_i\pr(M_i|D) \label{eq:AIC_cov} \\
      +&\sum_i\ev{a_0}_{M_i}\ev{a_0}_{M_i}^T\pr(M_i|D)
      -\qty(\sum_i\ev{a_0}_{M_i}\pr(M_i|D))\qty(\sum_i\ev{a_0}_{M_i}^T\pr(M_i|D)). \notag
\end{align}
The first term in \cref{eq:AIC_cov} can be computed with standard resampling methods, while the
other two give further contributions stemming from the spread of values the different models
produce.
We incorporate these last two terms in the form of Gaussian noise on the Bootstrap or Jackknife
distribution if we need the results for further calculations.
\Cref{fig:aic_example} shows an example for this model averaging procedure for the $\Omega$ baryon
on the N200 ensemble.

%% file: results.tex
\section{Results}\label{sec:results}

As the goal of this project is to find a suitable quantity for setting the lattice scale on CLS
ensembles with the inclusion of isospin breaking, we restrict our discussion on states which do not
decay in QCD+QED.
We summarize the currently achieved precision in \cref{tab:results_sym} for the asymptotic masses in
isospin-symmetric QCD.
In \cref{tab:results_ib} we focus on the $\Xi$ and $\Omega$ baryons which are promising candidates
for scale setting due to their precision and, in the case of the $\Omega$, its weak dependence on
the light-quark mass which results in a small isospin-breaking correction when compared to other
baryons.

\begin{table}[htb]
    \centering
    \caption{
        Relative errors $\frac{\Delta(am_i)\zth}{(am_i)\zth}$ of the asymptotic masses of all stable
        octet- and decuplet-baryons in the isospin-symmetric theory obtained from AIC model averaged
        single- and two-state-fits.
        All quoted values ignore systematic contributions to the error, but include the variation
        from the different fits going into the average as per \cref{eq:AIC_cov}.
    }
    \label{tab:results_sym}
    \input{tables/results_symmetric}

\end{table}
\begin{table}[htb]
    \centering
    \caption{
        Relative errors $\frac{\Delta(am_i)\fst}{(am_i)\fst}$ of the isospin-breaking corrections to
        the the masses of the $\Xi$ and $\Omega$ baryons from AIC model averaged single- and
        two-state-fits.
        All quoted values ignore systematic contributions to the error, as well as contributions
        coming from corrections to the sea-quark sector, but include the variation
        from the different fits going into the average as per \cref{eq:AIC_cov}.
    }
    \label{tab:results_ib}
    \input{tables/results_ib}
\end{table}

The precision quoted in these tables do not include systematic contributions to the uncertainty.
However, from the typical size of the statistical errors we find that we can push all of the
considered states below \SI{1}{\%} precision and, in the case of the $\Xi$, even below \SI{0.5}{\%}
for the isospin-symmetric masses on all ensembles.
The uncertainties in the isospin-breaking corrections are ususally $\mathcal O(\SI{1}{\%})$, which
we expect to be negligible when considering the full computation of the baryon masses in QCD+QED as
these corrections are multiplied by expansion coefficients of $\mathcal O(10^{-3}-10^{-2})$,
suppressing these uncertainties.
Since we do not include any corrections in the sea-quark sector, these uncertainties are likely
underestimated, but from the above argument, we would still expect the overall uncertainties to be
subdominant to the isospin-symmetric contribution's uncertainty.

%% file: tables/results_symmetric.tex
\begin{tabular}{|c||c|c|c|c|c|c|c|c|}
\hline
Ensemble & $N$ & $\Lambda$ & $\Sigma$ & $\Xi$ & $\Omega$ \\
\hline
D450 & 0.83\% & 0.67\% & 0.32\% & 0.28\% & 0.31\% \\
N200 & 0.97\% & 0.37\% & 0.37\% & 0.24\% & 0.33\% \\
N203 & 0.36\% & 0.28\% & 0.30\% & 0.24\% & 0.40\% \\
N451 & 0.25\% & 0.17\% & 0.15\% & 0.34\% & 0.20\% \\
N452 & 0.69\% & 0.38\% & 0.59\% & 0.24\% & 0.95\% \\
\hline
\end{tabular}

%% file: tables/results_ib.tex
\begin{tabular}{|c||c|c|c|c|c|c|c|c|}
\hline
\multirow{2}{*}{Ensemble} & \multicolumn{3}{c|}{$\Xi^0$} & \multicolumn{3}{c|}{$\Xi^-$} & \multicolumn{2}{c|}{$\Omega^-$} \\
\cline{2-9}
& $e^2$ & $\Delta m_u$ & $\Delta m_s$ & $e^2$ & $\Delta m_d$ & $\Delta m_s$ & $e^2$ & $\Delta m_s$ \\
\hline
D450 & 1.6\% & 2.2\% & 0.7\% & 0.9\% & 2.2\% & 0.7\% & 1.4\% & 1.7\% \\
N200 & 1.5\% & 1.8\% & 0.9\% & 1.0\% & 1.7\% & 0.9\% & 2.5\% & 2.6\% \\
N203 & 1.0\% & 1.2\% & 0.6\% & 0.8\% & 1.2\% & 0.6\% & 1.8\% & 1.8\% \\
N451 & 1.8\% & 1.4\% & 1.5\% & 1.2\% & 1.4\% & 1.6\% & 0.9\% & 1.0\% \\
N452 & 0.9\% & 1.0\% & 0.7\% & 0.6\% & 1.0\% & 0.7\% & 1.5\% & 2.4\% \\
\hline
\end{tabular}

%% file: conclusion.tex
\section{Conclusion and Outlook}\label{sec:conclusion}

We have presented the status of our investigation of baryon octet- and decuplet-masses as candidates
for scale setting on CLS $N_f=2+1$ ensembles with the inclusion of isospin-breaking effects.
Given the high statistical precision we observe in the isospin-symmetric mass determinations and the
relatively small size of isospin-breaking corrections, we are confident that we can achieve
sub-percent precision for the lattice scale.
Thus far, we only have data at two different lattice spacings of $\sim\!\!\SI{0.064}{fm}$ and
$\sim\!\!\SI{0.076}{fm}$, but we intend to add further ensembles, allowing for a preliminary
continuum extrapolation and better investigation of the precision we can achieve for the lattice
scales using these baryon masses.

%% file: acknowledgements.tex
\acknowledgments

The authors gratefully acknowledge the Gauss Centre for Supercomputing e.V. (www.gauss-centre.eu)
for funding this project by providing computing time through the John von Neumann Institute for
Computing (NIC) on the GCS Supercomputer JUWELS at Jülich Supercomputing Centre (JSC).